\documentclass[manuscript]{acmart}
\AtBeginDocument{%
  }

\begin{document}

\title{The Limits of AI Data Transparency Policy: Three Disclosure Fallacies}



\author{Judy Hanwen Shen}
\email{jhshen@stanford.edu}
\affiliation{%
  \institution{Stanford University}
  \city{Palo Alto}
  \state{CA}
  \country{USA}
}
\author{Ken Liu}
\affiliation{%
  \institution{Stanford University}
  \city{Palo Alto}
  \state{CA}
  \country{USA}
}

\author{Angelina Wang}
\affiliation{%
  \institution{Stanford University}
  \city{Palo Alto}
  \state{CA}
  \country{USA}
}

\author{Sarah H. Cen}
\affiliation{%
  \institution{Stanford University}
  \city{Palo Alto}
  \state{CA}
  \country{USA}
}

\author{Andy K. Zhang}
\affiliation{%
  \institution{Stanford University}
  \city{Palo Alto}
  \state{CA}
  \country{USA}
}

\author{Caroline Meinhardt}
\affiliation{%
  \institution{Stanford University}
  \city{Palo Alto}
  \state{CA}
  \country{USA}
}

\author{Daniel Zhang}
\affiliation{%
  \institution{Stanford University}
  \city{Palo Alto}
  \state{CA}
  \country{USA}
}

\author{Kevin Klyman}
\affiliation{%
  \institution{Stanford University}
  \city{Palo Alto}
  \state{CA}
  \country{USA}
}

\author{Rishi Bommasani}
\affiliation{%
  \institution{Stanford University}
  \city{Palo Alto}
  \state{CA}
  \country{USA}
}

\author{Daniel E. Ho}
\affiliation{%
  \institution{Stanford University}
  \city{Palo Alto}
  \state{CA}
  \country{USA}
}

\renewcommand{\shortauthors}{Shen et al.}

\begin{abstract}
Data transparency has emerged as a rallying cry for addressing concerns about AI—data quality, privacy, and copyright chief among them. Yet while these calls are crucial for accountability, current transparency policies often fall short of their intended aims. Similar to nutrition facts for food, policies aimed at nutrition facts for AI currently suffer from a limited consideration of research on effective disclosures. We offer an institutional perspective and identify three common fallacies in policy implementations of data disclosures for AI. First, many data transparency proposals exhibit a \textit{specification gap} between the stated goals of data transparency and the actual disclosures necessary to achieve such goals. Second, reform attempts exhibit a \textit{enforcement gap} between required disclosures on paper and enforcement to ensure compliance in fact. Third, policy proposals manifest an \textit{impact gap} between disclosed information and meaningful changes in developer practices and public understanding. Informed by the social science on transparency, our analysis identifies affirmative paths for transparency that are effective rather than merely symbolic. 
\end{abstract}

\received{20 February 2007}
\received[revised]{12 March 2009}
\received[accepted]{5 June 2009}

\maketitle

\section{Introduction}
Advocates for Generative AI transparency have made major strides in prioritizing transparency as a crucial mechanism for accountability among policymakers~\citep{AB2013,  bommasani2023foundation}. Similar to nutrition labels, policymakers want AI data transparency to provide clear, digestible information about AI systems. For example, California's AB 2013~\footnote{We provide a detailed analysis of the support for, opposition of, and compliance with AB2013 in Section~\ref{sec:ab2013}} requires developers to publicly post high-level summaries of datasets used in the development of generative AI systems or services~\citep{AB2013}. The EU AI Act similarly requires developers of general purpose AI models to disclose a data summary, including the types of data and whether any data is protected by copyright~\cite{EUAIAct}. These requirements arose in part due to impactful advocacy from the academic community for data disclosures as nutrition facts for AI~\citep{holland2018dataset, chmielinski2024clear}. However, like nutrition facts for food, the implementation of transparency objectives face their own set of policy challenges. Through an institutional perspective, we illustrate the critical misalignment that causes current policies to fall short of delivering their intended goals by identifying three fundamental gaps. 

The first fallacy is the \textit{specification gap} between the stated transparency goals, if they are even specified, and the necessary disclosures to achieve these goals. Although issues with the broad alignment of AI regulatory standards have been highlighted by previous work~\citep{guha2024ai}, the disconnect in data transparency policy, specifically between ideals and actual implementation, is pronounced. Current transparency requirements hint at several protections, yet the defined disclosures fall short of these aims, a topic seldom explored in the literature. Our work systematically identifies current and prospective transparency objectives and aligns them with minimum disclosure requirements.

The second fallacy is the \textit{enforcement gap} that exists between what companies are required to disclose and the mechanisms that ensure that the disclosures are made accurately. Given the invocation of nutrition labeling for AI policy, we first illustrate these enforcement challenges in US nutritional labeling itself. The transparency efforts in food regulation illustrate the many challenges that arise when considering the actors, incentives, power, and resources required to ensure meaningful transparency. These challenges are particularly acute for AI, where technical complexity further complicates verification efforts. 

The third fallacy is the \textit{outcome gap} between the data transparency compliance behavior and the model provider behavioral changes that are actually needed to protect individuals, data creators, and other stakeholders. Many works have highlighted the failures of mandatory disclosures~\citep{fung2007full, ben2017failure} and we identify how some of these failure modes also translate to AI transparency policy. Our contributions in this work are as follows:
\begin{itemize}
    \item \textbf{Survey motivations} for data transparency and connect data transparency objectives with the specific levels of data disclosures necessary (Table~\ref{tab:stakeholder-data-wide}), clarifying the \textit{specification gap}.
    \item \textbf{Illustrate the challenges} of enforcing data transparency requirements, drawing on social science and evidence from actual nutrition labels, showing the \textit{enforcement gap}.
    \item \textbf{Identify failure modes} of mandatory disclosures and how the \textit{outcome gap} impacts data transparency policy.
    \item \textbf{Outline a technical research agenda and policy refinements} to improve the robustness of transparency proposals.
\end{itemize}
Our work outlining these mechanisms and fallacies shows where technical work remains important (e.g., to address the specification gap) and where greater (non-technical) attention to implementation and institutional design is required. 
\section{Related Work}
Many calls have been made for understanding the training data of machine learning models, from documentation frameworks for datasets~\citep{guha2024ai} to audits of dataset licensing~\citep{longpre2023data} to Dataset Nutrition Labels aimed at avoiding incomplete or biased training data~\cite{holland2018dataset}. More information on the data used to train AI has been identified as a way to protect  public when AI is used in the public sector~\citep{thiel2023identifying, Adams_Greenwood_2024}. These works have motivated the emergence of disclosure requirements as a key regulatory measure in the age of foundation models \citep{anderljung2023frontierairegulationmanaging, 10.1145/3593013.3594079}. Legislation such as California Assembly Bill 2013 (AB2013) on Generative Artificial Intelligence: Training Data Transparency for System Developers (2024) and Regulation (EU) 2024/1689 of the European Parliament and of the Council on Artificial Intelligence (the EU AI Act) include multiple types of disclosure requirements for system developers and model providers, respectively. This body of work has been instrumental in establishing data transparency as a cornerstone of AI accountability.

Building on these contributions, scholars have begun examining whether transparency requirements as implemented achieve their intended aims. \citeauthor{guha2024ai} highlight the disconnect between AI regulation and its intended purpose, \textit{the regulatory alignment problem}, arguing that disclosure requirements in AI regulation may be feasible and inexpensive to adopt, but may not result in actionable protection of consumers. 
With respect to data transparency for foundation models, the Foundation Model Transparency Index~\citep{bommasani2023foundation} introduces 10 data-related indicators motivated by the role of data disclosures for corporate accountability. These indicators include whether developers disclose information on data size, composition, processing, and curation, as well as whether copyrighted, proprietary, or personal data is included. 
An index approach allows a comparison of the transparency practices of various model developers, but the specific desired impact (e.g. copyright protection, privacy protection) of improved transparency for each indicator is unclear. \citeauthor{ananny2018seeing} interrogate transparency as an ideal and how transparency for the sake of transparency may not be an effective tool to achieve algorithmic accountability. The presence of disclosures does not necessarily guarantee adequate auditing to protect users, content creators, and data workers. In a grounded analysis of 18 targeted transparency policies, ~\citeauthor{fung2007full} highlight how public disclosures are frequently ineffective and counterproductive due to information quality and loopholes found by disclosures'; targeted disclosures must be user-centric and sustainable to be effective. \citeauthor{ben2017failure} point out that most people do not find the flood of information provided through mandated disclosures comprehensible nor useful, with \citeauthor{zalnieriute2021transparency} adding that \textit{transparency-washing} can serve the interests of major technology companies by providing the illusion of substantive change. Many critiques of transparency for algorithmic accountability have highlighted that transparency initiatives are too limited in their aims \citep{10.1145/3593013.3594104}, do not produce substantive change without grassroots pressure \citep{boyd2016algorithmic}, and are easily gamed \citep{doi:10.1177/20539517231164119}. 

Gaps in the AI regulatory landscape such as environmental sustainability~\citep{hacker2023sustainable}, AI for health care~\citep{palaniappan2024gaps}, and AI transparency broadly~\citep{howell2024policy} have also been highlighted. For example, \citeauthor{hacker2023sustainable} identifies how European legislation intended to govern the environmental impacts of AI needs to be improved and emphasizes that policy must go beyond transparency mandates. \citeauthor{howell2024policy} examines the \emph{interoperability} challenges across different jurisdictions and highlights the need for more details on AI transparency requirements in general. Furthermore, the policy landscape of data transparency may be difficult to navigate. In a meta-review of 2200 papers related to trustworthy AI, \citeauthor{micheli2023landscape} find that there are gaps between existing methods of documenting datasets and European regulatory objectives.  

Our work follows a similar path by first identifying critical gaps in implementation of AI disclosures in order to then suggest policy remedies that best help achieve the actual goals of AI transparency. Specifically, we add to this discussion by focusing on data transparency (in comparison to, for instance, model transparency) from an institutional perspective. By focusing on data transparency, we can provide a deep analysis of forms of data transparency alongside various policy objectives, with a particular eye on data-specific enforcement challenges.
\section{Levels of Data Transparency}
\label{sec:levels}
Despite data transparency being a frequent goal in AI policy, \textit{transparency} can refer to different disclosures \citep{}. We taxonomize different forms of data transparency based on the artifact that the disclosure process produces; using this taxonomy allows a deeper analysis of the burden of providing these disclosures and the technical feasibility of validating these disclosures. This section provides a common transparency language around which we develop our later discussions. For each transparency requirement, we reference AB2013, the EU AI Act (including both requirements for General-Purpose and High-Risk) AI systems), and GDPR examples for where this requirement appears. The purpose of this section is to take inventory of the types of existing and potential data transparency rather than to provide a deep legal analysis of existing disclosure requirements. 

\subsection{Level 1: Documentation and Descriptions} 
This level of transparency includes various information \emph{surrounding} the datasets used in the development of generative AI models. We further separate data information into the following two categories: information about the dataset itself (e.g., how data is processed and used) and information about the dataset supply chain (e.g., how data is acquired). This level of disclosure produces an artifact that is a documentation of the data.  

\textbf{Level 1a: Dataset Information}
Dataset information is metadata on the content and construction of the dataset. Some examples, along with those for which existing regulations mandate them, include
\begin{itemize}
    \item Summary Statistics (AB 2013): High-level dataset statistics, such as the number of training points.
    \item Data Sources (AB 2013): Sources (e.g. web, books, video) from which training data was gathered and what purpose each source serves.
    \item Synthetic Data (AB 2013): Description of whether and how synthetic data was used in the training data, and how the synthetic data was generated. 
    \item Personal Information (AB 2013, EU AI Act): Whether the datasets used contain personally identifiable information or aggregate consumer information.
    \item Copyrighted Content (AB 2013): Whether the datasets contain copyright, trademarked, or patented information or whether the datasets are in the public domain. 
    \item Data Processing (GDPR, AB 2013, EU AI Act): Cleaning, filtering, removal of PII and other processing steps taken before the data were used for model training.\footnote{GDPR's requirements apply to the data processor which may be different from the model provider in some settings.} 
\end{itemize} 

Verifying this level of information though observing only a trained model is extremely challenging. Namely, a model can ``memorize'' a piece of content without ``regurgitation''~\citep{cooper2024files}. This distinction is important because it means that even when personal information or copyrighted content is a part of training data, generating that content may be hard. Even when the goal is ``extraction'', when adversarial users intentionally try to make a model produce a certain output verbatim, not every piece of data in the training dataset can be extracted successfully~\citep{nasr2023scalable}. Furthermore, overlap in content between data sources makes extraction and membership inference difficult. 

\textbf{Level 1b: Data Supply Chain Information} 
We consider the data supply chain broadly to include many different types of data, including public datasets, publicly available data, licensed data, human-generated data, and synthetically generated data. Here, data supply chain refers to the process and network by which data used for AI is produced~\citep{hopkins2025ai, lee2023talkin, widder2023dislocated, cobbe2023understanding}. This level of information around the data-generating process of the datasets used involves both the data producers and the model developers. Although existing regulations focus on system developers and model providers, it is likely that these parties know or control aspects of the supply chain when they work with data producers. In this type of disclosure, documentation and descriptions include the following: 

\begin{itemize}
    \item Data Collection (AB 2013, EU AI Act): Information about when, how, and from whom the dataset was collected.  
    \item Consent (GDPR): If training data includes personal data, whether and how consent was collected. 
    \item Licensing (AB 2013): Information about whether the dataset was purchased or licensed; permissions around dataset usage. 
    \item Vendors: Information on which data vendors were used in the dataset collection, labeling, and processing.   
    \item Contracts: Contracts between data providers and model developers that allow access to data for training and evaluation. 
    \item Data Workers: Details about who labeled and collected the data and how they were compensated or attributed. 
    \item Data Provenance: Information on how the dataset is derived including sourcing, creation, and licensing.\footnote{There is overlap with the prior bullet points, but the focus is on understanding the original sources which training datasets stem from~\citep{longpre2023data}.}
\end{itemize}

Similar to Dataset Information (1a), verifying this category of information is difficult without access to the entire training dataset. For example, from querying a model, it is impossible to discern whether consent was collected from individuals before their data were used for training or whether the reported compensation given to data workers was accurately reported. Even the push to verify these disclosures may need to come from complaints from consumers, data workers, or data vendors who know that the reported information is inaccurate.  

\subsection{Level 2: Data Access}
This second level of transparency involves direct access to training data in some form. Although not traditionally considered a \textit{disclosure}, access to training data can be required to achieve certain policy goals, despite being costly or in contradiction to the business interests of model developers. 
For proprietary models, datasets are often secret, and dataset descriptions are not equivalent to direct access to datasets. When data contain personal information, access to anonymized data also falls under this category of disclosure, since anonymized data can be included as a subset or as a full dataset. Access to a dataset itself is not binary; we discuss three forms: membership access, subset access, and full data access.

\textbf{Level 2a: Membership Access} Given a document, such as a webpage, a book, or a code repository, a membership query access returns a yes or no answer about whether the piece of data is included in the dataset. The technical feasibility of such an access is contingent on the length and uniqueness of the data segment since overlap itself is a fuzzy concept. Relaxations of this notion include searching for parts of a document through keywords or phrases or approximate matches to the query document. 
 
\textbf{Level 2b: Subset Access} Partial access to training data or the generation of training data (e.g., public datasets, prompts) (EU DSA) includes when portions of the dataset, transformations of public datasets, or instructions for generation of datasets (e.g., prompts) are provided. Partial access also includes access to a limited set of data samples. To illustrate the format of the data used in different phases of training, some samples can be provided on the type of data used for each phase of training. Although this gives insight into the real data used, it does not reveal all of the data choices. 

\textbf{Level 2c: Full Data Access} The entire dataset used for one or more components of training, validation, and testing of the model before deployment is made available. This includes new data used after substantial changes to the model. Full dataset access would also imply access to the mentioned transparency levels (e.g., 1a, 2a, and 2b).    

When verifying the correctness of each of these forms of data disclosure, it is important to consider whether the disclosed samples are actually used (precision) and whether the disclosed data are not exhaustive of all the training data used (recall). Most works focus on recall in the setting where training data points are available; data attribution methods can be used to test whether training datasets were not reported~\citep{park2023trak, grosse2023studying, worledge2024unifying}. However, without knowledge of what pieces of data to test for, verifying recall (e.g., all training data were reported) is very difficult. 
Methods of verifying that the data provided in levels 2b and 2c were indeed used in the training set include comparing the likelihood of the training data vs data not included in training~\citep{shokri2016membership}. 

\begin{table*}[htbp]
\centering \small 
\setlength{\tabcolsep}{4pt} 
\renewcommand{\arraystretch}{1.3} 
\begin{tabular}{p{0.25\textwidth}p{0.27\textwidth}p{0.40\textwidth}}

\textsc{Subobjective} &  \textsc{Minimum Disclosure} & \textsc{Verification} \\
\hline
\multicolumn{3}{|l|}{\textbf{Protecting Personal Information}} \\
\hline
Understand the presence of PII of citizens in training data & (1a) Dataset Information: Personal Information \newline (1b) Data Supply Chain Information: Data Collection & Regulators may need access to the training dataset to verify or rely on complaints of models revealing personal information. \\
\hline
Choose AI products and services that do not use personal data & (1a) Dataset Information: Personal Information & Individual users do not have the tools to test comprehensively whether personal data has been used \\
\hline
Discern whether their name and likeness were included in the training data for a specific mode & (2a) Membership Access: Membership query access to training data & Users can test whether their information can be revealed by the model, but cannot verify that their personal data was not used at all~\citep{cooper2024files}. \\
\hline
\multicolumn{3}{|l|}{\textbf{Assurance of Training Data Quality}} \\
\hline
Assurance of diverse data collection procedures & 
(1a): Dataset Information: Data Sources
(1b): Data Supply Chain Information: Data Collection & 
Regulators can verify that data sources are diverse by higher levels of data access (e.g., 2a membership access) \\ \hline 
Assurance of data quality & 
(1a): Dataset Information: Data Processing
(1b): Data Supply Chain Information: Data Collection & 
Regulators can verify that data processing has been done to improve quality by inspecting source code for data.   \\ \hline 
Interpret whether predictions can be trusted for their specific individual profile & 
(1a): Dataset Information: Data Sources
(1b): Data Supply Chain Information: Data Collection & 
Individual users could verify through higher levels of data access (e.g., 2b subset access) \\ \hline 
Assurance of Data Representativeness & (2b): Subset Access 
(2c): Full Data Access & \\ \hline 
\multicolumn{3}{|l|}{\textbf{Copyright and Terms of Use Protections}} \\
\hline 
Assurance of copyright protection & 
(2a) Membership Access: Membership query access to training data & 
Verifying correctness of membership query responses relies on access to the actual training data. \\ \hline
Assurance of copyright and licensing law compliance &
(1b) Data Supply Chain - Licensing & 
It is difficult for regulators to verify that all data used is licensed. Regulators may need to rely on complaints from data owners. \\ \hline 
Compliance with terms of use & 
(1b) Data Supply Chain - Data Collection & 
Model developers may verify that competitors did not train on data generated by their models through watermarking their outputs.~\citep{kirchenbauer2023watermark} \\ \hline
\multicolumn{3}{|l|}{\textbf{Evaluation Generalization}} \\
\hline 
Assurance of no train-test overlap & 
(1a) Dataset Information: Data Processing & 
While it is impossible to verify the entire data processing pipeline, it might be possible to identify significant omissions through model behavior~\citep{golchin2023time, Shi2023DetectingPD}. \\ \hline
Check for the presence of evaluation examples in the training data & 
(2a) Membership Access: Membership query access to training data & 
Verifying that membership query responses requires access to training data. However, it may be possible to observe behavior on benchmarks to infer potential contamination~\cite{zhang2024language}.\\ \hline
\multicolumn{3}{|l|}{\textbf{Data Laborer Protections}} \\ \hline 
Choose AI platforms and services that are produced via fair compensation & 
(1b) Data Supply Chain Information: Data workers & 
Currently consumers cannot verify that AI platforms fairly compensated data workers. \\ \hline 
Check that forced or child labor is not a part of generating data & 
(1b) Data Supply Chain Information: Data workers & 
Regulators may use a complaint system to censure companies that engage in labor practices they are not reporting. \\ \hline 
\end{tabular}
\caption{Overview of mapping between the objectives of data transparency and the minimum level of \textit{data} disclosure. Different disclosures suffer from different challenges in verification.}
\label{tab:stakeholder-data-wide}
\end{table*}

\section{The Specification Gap}
We now map each data transparency objective to the requisite levels of data transparency that would enable the objective. In doing so, we illustrate the specification gap between mandated and required disclosures for each objective. For example, if the goal is to assess copyright infringement, what is the actual form of transparency that allows this assessment? 
Table~\ref{tab:stakeholder-data-wide} provides a summary of the fine-grained sub-objectives and the necessary disclosures and the corresponding verification requirements.\footnote{The corresponding data disclosure levels are described in Section~\ref{sec:levels}.} Ultimately, there are two paths forward, either through increasing the level of mandated disclosures or through technical advancement that can provide verification through less revealing disclosures. 

\subsection{Protecting Personal Information}
\label{sec:personal-info}
Data transparency around personal information is often for protecting individuals and their data. This type of data transparency could empower consumers to make informed decisions around AI products and services that use personal data for training. Data transparency alone is not enough to protect these rights since downstream actions based on privacy law is required. However, data transparency provides the necessary first step of disclosure of when personal information is used. 

\textbf{Mandated Disclosures}
The European Union’s regulatory framework, specifically the General Data Protection Regulation (GDPR) and the EU AI Act, establishes mandatory disclosure requirements for the processing of personal data. Under Article 13 of GDPR, organizations must explicitly inform data subjects about the intended purposes of data collection at the time such data is obtained. Furthermore, Article 10 of the EU AI Act imposes additional obligations on providers of high-risk AI systems to implement data governance practices that protect `individuals' fundamental rights and freedoms. In California,  AB 2013 § 3111(a)(7) specifies that AI system developers must provide information about whether the datasets include personal information or aggregate consumer information. 

\textbf{Necessary Disclosures} Although these provisions aim to protect users if their data are collected for training AI systems, the actual disclosures required for these protections depend on the goal. Current data transparency policies mandate only Dataset and Supply Chain disclosures (1a, 1b), which enable high-level understanding of PII presence and consumer choice. However, these are insufficient for users who want to verify whether their personal information was actually used in training; Membership Access (2a), which no current policies mandate, is required (Table~\ref{tab:stakeholder-data-wide}). Because providing such access is costly and may reveal proprietary training pipelines, technical advances in membership inference methods are important for privacy verification without direct access (e.g., through model behavior).\footnote{See Section~\ref{app:research-dir} for detailed recommendations for research directions.}

\subsection{Assurance of Training Data Quality}
Existing regulation around training data quality is driven by the goal of reducing harm or discrimination experienced by downstream users. In the computer science literature, the importance of data quality has been highlighted~\citep{xu2021detoxifying, longpre2024pretrainer, wettig2024qurating}; sometimes as a factor even more important than the size of the dataset~\citep{zhou2023lima, shen2024data}. 

\textbf{Mandated Disclosures}
For high-risk AI systems (e.g., health care, criminal justice), the EU AI Act specifies requirements for training, validation, and testing data. In Article 10(3) of the EU AI Act, datasets are required to be “sufficiently representative, and to the best extent possible, free of errors and complete because of the intended purpose”. This requirement is difficult albeit not impossible because of language like ``sufficiently'' and ``to the best extent possible''. Other requirements for high-risk data include disclosure of datasets used to deployers (Article 13(3)(b)(vi)) and documentation of data collection and processing details (Article 17(1)(f)). 

\textbf{Necessary Disclosures}
For regulators and individuals interested in steps taken to improve data quality, level 1 disclosures are sufficient since the inclusion of various data sources can be reported through these disclosures. If an auditor wanted to evaluate the degree to which the training data is representative of a customer segment or society more generally, a sufficiently large random sample (2b) or full data access (2c) would be necessary. Thus, the full validation of training data quality often requires levels of data transparency that existing policy does not mandate. 

In some cases, the actual goal of improving data quality might be to ensure that the final model does not exhibit undesirable behaviors (e.g., biases, dangerous answers). However, decisions can be made during model training to mitigate undesirable behavior, even if the training dataset is flawed. For example, even if a dataset does not reflect the distribution of users using the downstream product, data points representing minority views or preferences can be up-weighted to mitigate the downstream biases of a model. This kind of reweighing can happen through data pre-processing (Dataset Information 1a) or through algorithm design, which no transparency requirements currently cover. Thus, data transparency alone is not effective since a representative dataset is neither sufficient nor necessary to achieve an unbiased downstream model. 

\subsection{Copyright and Terms of Use Protections}
\label{sec:copyright}
As generative AI becomes increasingly capable of producing high-quality creative and editorial content, concerns have been raised about whether the data used to train these models contain copyrighted materials. Copyright is assigned from the moment a piece is created~\citep{USC17_102, BerneConvention}, and thus the massive scale of datasets of books and images that are used to train generative models often include copyrighted material~\citep{ bandy2021addressing, karamolegkou2023copyright}. However, the enforcement of copyright protections often falls in the hands of the creators. As a result, supply chain transparency~\citep{lee2023talkin}, including filtering of training data to reduce the likelihood of copyright infringement~\citep{henderson2023foundation}, has been proposed. Data transparency disclosure requirements is again only a necessary but not sufficient step for copyright protections. The topic of fair use for generative models is still under debate; if training on copyrighted content is fair use, data transparency would not protect creators. 

Furthermore, model creators and developers often require the usage of their services to comply with their terms of use (TOS). For example, OpenAI's terms of use specify that model output cannot be used for the development of models that compete with the company.\footnote{\url{https://openai.com/policies/row-terms-of-use/}} At the same time, the terms of use grant users ownership of the generated content. \citeauthor{lemley2024mirage} highlight that TOS restrictions face legal challenges since it is unclear whether copyright applies to model weights and outputs. However, requirements for disclosing and specifying how data is collected or created may disincentivize companies from generating data using other platforms that explicitly prohibit this practice. 

\textbf{Mandated Disclosures}
In European markets, the EU AI Act specifies that “Providers of general-purpose AI models shall: put in place a policy to comply with Union law on copyright and related rights” (Article 53(1)(c)). Although no transparency requirement is explicitly specified, compliance may require model providers to confirm that copyright data filtering was performed as part of the training data filtering process. For noncommercial purposes, in particular, there are exemptions for training models using copyrighted data. AB2013 specifies that training data transparency should include § 3111(a)(5-6) “Whether the datasets include any data protected by copyright, trademark, or patent, or whether the datasets are entirely in the public domain. Whether the datasets were purchased or licensed by the developer.” By directly requiring the transparency of training data, creators can theoretically sidestep the tricky legal landscape of eliciting the outputs of generative models. 

\textbf{Necessary Disclosures}
While disclosing whether copyright data was used in training data seems low-cost, these very same disclosures are too vague to be useful for creators. Even if companies claim that no copyright data was used, it would be difficult to verify these claims for all copyright-related content. The more effective level of disclosure would be to provide access to the data creator in the membership access (2a) (Table~\ref{tab:stakeholder-data-wide}). 
At the data collection and curation step, upstream in model training, revealing data sources and their associated licensing may be helpful as an additional incentive to protect copyright. Meanwhile, the regulatory landscape of AI copyright continues to shift rapidly as creators are increasingly using AI to assist with content creation~\citep{hristov2016artificial, usco2025ai}. This uncertainty creates a moving target reduces the effectiveness of documentation and description (1a, 1b) disclosures since the notion of what is copyright content keeps evolving.

\subsection{Forward-Looking Objectives}
Next, we look beyond existing regulation to understand the potential future objectives of transparency for training data used for generative AI. Although these objectives have been discussed for machine learning in general, translating these objectives to generative AI models has not yet appeared in legislation.\footnote{We also include 3 additional objectives forward-looking objectives in Section~\ref{app:other-objectives}.} 

\textbf{Evaluation Generalization} Model evaluations are only valid if there are guarantees evaluation examples are not in the training data or in-context examples. This is often called ``train-test overlap''~\citep{kapoor2022leakage, golchin2023time, zhang2024language}. Data transparency could theoretically ensure the validity of evaluations against standard benchmarks. The closest existing policy is the EU AI Act Article 10(3) specifying that training data to be free from errors as much as possible. Ensuring that training data do not contain evaluation data could fit under this requirement if evaluation data is considered an erroneous inclusion. However, verifying test set contamination is difficult and more technical research is required. Recent work has shown that specific passages can be generated verbatim even without the passage appearing in the training data~\citep{liu2025language}. 
Furthermore, deleting information from models through methods such as unlearning does not provide guarantees about changes in model output~\citep{cooper2024machine}. Finer grained notions of data usage, efficient membership tests, and cryptographic approaches to data usage (Section \ref{app:research-dir}) would help minimize the necessary policy disclosures for preventing benchmark contamination.

\textbf{Data Laborer Protections}
Many works have shown the importance of data quality in fine-tuning generative models~\citep{zhou2023lima}. Existing legislation requires supply chain transparency to protect the labor that generates high quality data. For example, the California Transparency in Supply Chains Act, Cal. Civ. Code § 1714.43 (2010) was designed to help consumers make informed decisions by requiring retail sellers and manufacturers to disclose company standards for trafficking and slavery in their supply chain. As another example, the Uyghur Forced Labor Prevention Act, Public Law No. 117-78, 22 U.S.C. § 6901 prevents the import of goods made wholly or in part by forced labor in the People’s Republic of China. 
Recent investigative reports on data labor practices of large AI companies found that Kenyan workers who provide data annotations are tasked with traumatizing work without ensuring job security~\citep{perrigo2023chatgptkenya}. Protections proposed for data workers have been proposed but rarely adopted~\citep{pai2024protecting}. AI transparency policy should explicitly require disclosures that include detailed information about the AI \textit{ghost work}~\citep{gray2019ghost} behind training data in order to better protect workers.

\section{The Enforcement Gap}
\label{sec:remediation}
Nutrition Facts have been a foundational motivation for AI transparency ~\citep{holland2018dataset, gebru2018datasheets}, but this archetype also reveals how common AI policy implementations have failed to wrestle with the institutional reality of nutrition labeling. The experience of US nutrition labeling -- a story of delegation to private companies with critiques of abdication of public enforcement -- typifies what we  call the enforcement gap, offering important lessons for AI proposals. 

\subsection{The Archetype of Nutrition Facts}
Nutrition facts first became widespread in the United States in 1973 when the FDA proposed standardizing nutrition labels on foods~\citep{wartella2010history}. This effort was in response to consumers' requests that better information about food be available with the rise of processed foods. By providing nutrition facts, food producers theoretically enable consumers to better choose what they consume. Despite these well-motivated objectives, it is subject to serious critique \citep{heinzerling2015varieties}. As Heinzerling argues, the ``existing legal system for food fails to deliver the transparency it seems to promise'' -- absent enforcement, formal disclosure amounts to obscurity in fact.  

\textbf{Public Enforcement:} The first challenge is regulatory fragmentation and capacity. In the U.S., the USDA and FDA, the two principal agencies in charge of food labeling, have different processes for reviewing food labels and different definitions of food claims. Labeling emanates from the Food, Drug, and Cosmetics Act, the Federal Meat Inspection Act, the Poultry Products Inspection Act, the Egg Products Inspection Act, the
Agricultural Marketing Act, and the Fair Packaging and Labeling Act. To compound the issue, the FTC has regulatory jurisdiction over deceptive advertising and associated claims. Despite the number of federal authorities, resources are limited. The capacity for public enforcement is dwarfed by hundreds of thousands of claims in the marketplace. 
From 1998 to 2008, the FDA secured only two court injunctions for misleading labeling~\citep{heinzerling2015varieties}. Despite the fact that FDA found in random samples in 1994 that 48\% of products misrepresented vitamin A and C volumes, both USDA and FDA abandoned attempts at verification via random sampling (id.). Food labeling, then, has been functionally delegated to private companies. 

\textbf{Private Enforcement}: 
Perhaps private enforcement could make up for the lack of public enforcement. But there is no federal private cause of action in the United States for consumers to sue. The Lanham Act establishes private right of action for \emph{competitors} to seek remedies when harmed by deceptive representations. POM Wonderful, for instance, successfully sued Coca-Cola for misleading claims of using pomegranate juice~\citep{pomwonderful2014}, even whilst their own claims of the disease-preventive properties of pomegranate juice were investigated by the FTC~\citep{pomwonderful2015}. Consumer groups have instead pursued state-level claims, such as under California's Unfair Competition Law. But such suits are resource-intensive and the effect on the reliability of claims overall has been limited. Despite the systems goals of transparency, the effect may be outright confusion. As stated by one former FDA Commissioner, food labels are ``so opaque or confusing that only consumers with the hermeneutic abilities of a Talmudic scholar can peel back the encoded layers of meaning. That is because labels spring not from disinterested scientific reasoning but from lobbying, negotiation, and compromise.''~\citep{stark2012ineffective} 

Put differently, far from a paragon for AI transparency, nutrition labeling illustrates the remedial gap: the inability to secure remedies for inaccurate, deceptive, or unreliable disclosures. Transparency is only useful insofar as it can be effectively audited and enforced. Comparative evidence shows that labeling may be designed in more effective ways. Food policies in Mexico, for instance, show signs of stricter enforcement~\citep{crosbie2023implementing, mexiconewsdailyProfecoWithdraws}. 
With this understanding in mind, we now assess the implications for AI transparency in two ways: (1) Compliance with Mandates: Do companies share data information as directed by the data transparency policy? and (2) Verifiability of Disclosed Information: Is the information shared by the companies verifiably correct?

\subsection{Component 1: Compliance with Transparency Mandates}
As bills like California’s AB 2013 and the EU AI Act mandate data information disclosure, the first goal of enforcement is to require model developers to comply with necessary disclosures. Their scope differs in that AB2013 requires only generative AI models to comply, while the EU AI Act applies to all AI systems and provides different requirements for transparency depending on a system's risk tier. However, current enforcement mechanisms for these disclosure requirements are weak and may not promote full compliance.

\textbf{Public Enforcement}: The EU AI Act puts forth both transparency requirements and specific guidelines for making disclosures public. For example, in July of 2025, the Commission released guidelines for summaries that providers of General-purpose AI models need to disclose. These guidelines include: 
\begin{itemize}
    \item \textbf{What} General-Purpose AI providers needed to disclose to the public through a template, 
    \item \textbf{How} providers should disclose this information (through their company website), 
    \item \textbf{When} these disclosures need to be available by (August 2, 2025). 
\end{itemize}
If these guidelines are not met, model providers are subject to fines administered by the commission starting in August 2026). In contrast, it is not clear what bodies could be responsible for enforcing data transparency requirements in the US. First, AB2013 does not assign a specific agency with enforcement and does not discuss auditing and verification. Enforcement will likely take place under California's Unfair Competition Law, which allows the California Attorney General's office to bring public enforcement actions. Such enforcement would be resource intensive, when there is little expertise within state government around AI issues. 
Second, AB2013 makes monitoring exceptionally difficult, as there is no standardization for data disclosures are important. AB2013 requires only that data transparency statistics, in some form, be posted on the developer’s website, and California's experience with privacy disclosures, which similarly lacked any standardization, were decried as ``functionally useless.''~\citep{luthi2021functionally} The California Privacy Protection Agency (CPPA) has brought actions against data brokers for failing to register and pay an annual fee to fund the California Data Broker Registry, which hosts a data deletion mechanism. However, these legal actions can be costly and may not succeed.  
Third, while noncompliance with the transparency provisions Article 50 of the EU AI act subjects model developers to fines up to €15,000,000 or up to 3\% of annual worldwide turnover (Article 99), California UCL violations are subject to civil penalties of up to \$2500 per violation.

Securing compliance with disclosure can also lead to unintended consequences, by siphoning enforcement resources from more nefarious practices ~\citep{ho2012fudging}. To avoid this type of outcome, both disclosures and subsequent compliance mechanisms must be designed taking into account the implementation burden.

\textbf{Private Enforcement}: While the EU AI Act does not create an explicit private right of action for individuals~\citep{evans2024euaiact}, California law provides a path for private enforcement of AB2013 under California's Unfair Competition Law. However, this path remains limited. First, private parties must suffer injury and financial loss because of an actor's failure to abide by the disclosure requirement of AB2013. A party who owns copyright may not be able to assert harm because a developer trained on copyrighted data, but must instead point to harm for the failure to \emph{disclose} intellectual property status. 
Second, private parties may not seek civil penalties, and can seek only injunctive relief or restitution. Private litigants hence have greater incentive to pursue other paths for recovery, such as  copyright or privacy law. 
Abiding by AB2013, however, can increase the risk of litigation, providing a powerful disincentive for developers to fully disclose~\citep{longpre2024dataaut}. Developers may rationally decide not to comply with transparency requirements -- given the lower risks of non-compliance -- than expose themselves to more serious liability -- including fines and criminal sanctions -- under copyright or privacy law~\citep{lee2023talkin}. 

\textbf{Takeaways:} Unlike the EU AI Act, there is no clear path to enforcement under AB2013 (See Section~\ref{sec:ab2013}). At a minimum, the enforcement of transparency compliance should be based on three principles. First, legislation should clearly allocate enforcement responsibility to a primary agency. Second, that agency should have adequate resources to carry out necessary enforcement actions. Third, penalties should be proportional to the scope and scale of violations. On each of these counts, AB2013 commits the same sin of nutrition disclosure: high aspirations coupled with low enforcement. 

\subsection{Component 2: Verifiability of Data Transparency Disclosures}
Once disclosures are made available, the second component of enforcement is to ensure that the disclosures are accurate. In some food label cases, competitor companies took on the burden of verifying or disproving false claims made by some companies~\cite{pomwonderful2014}. Although verifying the accuracy of food labels is laborious, the feasibility of verified data disclosures for AI depends on the even evolving technical landscape. Yet verifiability is essential since truthful data transparency enables the intended objectives of policymakers. 
Verifiable data transparency gives evidence to pursue legal action in areas such as copyright, privacy, and discrimination. 

\textbf{Public Enforcement}: Although centralized enforcement may allow a specific agency to have adequate authority to ensure that companies make necessary disclosures, verifying that disclosures are accurate requires significant domain expertise. Investigating complaints that data disclosures are misleading is likely beyond the capacity of existing agencies in charge of privacy or consumer protection due to a lack of AI expertise~\citep{jurowetzki2025private}. In nutrition, the FDA and USDA are agencies staffed largely by expert scientists, and yet verification of a significant number of food products is already too costly. For data transparency, government agencies have few if any AI experts. 

\textbf{Private Enforcement}: The success of legal actions for copyright, privacy, and anti-discrimination depends on verified claims of improper data usage. Verifying whether data disclosures are accurate may become the burden of plaintiffs seeking to sue model developers. For example, harm due to misrepresenting dataset content may give competitors incentives to challenge under California's Unfair Competition Law. 

\textbf{Technical Feasibility}: As highlighted in the right column of Table 1, the high-level statistics outlined in AB2013, such as the range of the number of training points and whether copyrighted and personal information is used, are too general to verify with only model access. Currently, even verifying the existence of a specific data source or data point used for training in enormous datasets can be difficult (see the Research Directions Section). 

\textbf{Takeaways:} Verifying data disclosures is difficult, but necessary to ensure the disclosures are truthful. Data disclosure requirements should (1) create systems, authorities, and infrastructure that enable verification of disclosures, and (2) not require disclosures that are impossible to verify, without creating adequate supporting processes. However, since significant technical barriers still remain to verified data disclosures through model access, we point to directions of technical research needed towards this effort of verifying disclosures in a later section. The status of the research there will determine what kinds of policy are enforceable in practice.

\section{The Impact Gap}
\label{sec:outcome-gap}
Even when mandated disclosures provide sufficient information with adequate enforcement, there is still a gap between intended goals and actual impact. For data transparency policies to achieve accountability for AI system developers, including the protection of personal information from consumers and the copyright of creators, the right mechanisms for impact must be in place. The failures of mandated disclosures across domains from privacy regulation to vehicle safety ratings have been well documented~\citep{ben2017failure}. Three key challenges that contribute to this impact gap include: whether the information released is truly informative for consumers and businesses; whether there are viable alternatives in the market; and whether these factors effectively incentivize companies to change their behavior in a way that aligns with the original policy objectives. These challenges particularly plague disclosures of AI data transparency. 

First, the impact of disclosures depends on whether the audience of the disclosure can adequately understand them. In federal home loans, for example, the 50 different disclosure forms overwhelm consumers, making it difficult to comprehend the risks that are being disclosed, and therefore these disclosures themselves do not curb predatory lending~\citep{stark2012ineffective}. Privacy policies are another common example of overwhelming consumers with information that they cannot possibly absorb~\citep{oeldorf2019overwhelming, o2024no}. For AI systems, complete disclosures including data sourcing, collection, and processing would create the same information overload for average consumers. Instead, data transparency mandates should focus on enabling information intermediaries such as journalists, academics, and nonprofits to translate these complex disclosures into actionable insights for consumers. 

Second, bad data practices may be ubiquitous in model training, limiting the effectiveness of disclosures in driving market-based accountability. Modern generative AI models are often pre-trained on as much of the Internet as possible, with companies mostly adopting different data approaches to achieve nuanced goals like safety and preference alignment. This ubiquitous `throw everything in' approach for pre-training severely limits consumer optionality. For example, it is possible that all models good at news writing need to be trained with copyrighted content. For transparency mandates to successfully generate consumer pressure on AI system developers, meaningful alternatives must be available, a condition that current industry practices make difficult to satisfy. 

Third, consumer pressure must incentivize the desired corporate behavior~\cite{wang2024strategies}: better practices around data provenance and quality. Companies might respond to transparency mandates by creating technically compliant but low-quality disclosures, such as listing thousands of data sources without meaningful prioritization or using vague categorical descriptions like `publicly available online content'. 
Returning to the example of food labels, the disclosure of sesame allergens has led to increased reports of sesame as an ingredient, as there is a lower likelihood of any penalty for over-reporting~\citep{apnewsLabelUnintended}. Similarly, model developers can relabel, reprocess, or cherry pick datasets rather than reducing the use of sensitive data. \citeauthor{fung2007full} highlight that effective disclosures give information to consumers that changes their behavior, which in turn changes the behavior of disclosers in a way that serves the original policy's goals. As AI advances rapidly, policymakers must iterate on data transparency policy to address evasive tactics when they arise.
\section{Recommendations}
For transparency policy to be well-specified, either disclosures must become more extensive (facing considerable barriers) or technical advances must bridge the gap between what is disclosed and what can be reliably verified. Beyond specification, effective implementation is also necessary because policies must be enforceable and create meaningful paths to impact. We conclude with related recommendations for technical research and policy design.

\subsection{Research Directions for Computer Scientists}
\label{sec:research-dir}
Techniques for verification are important to enable weaker disclosure requirements to achieve meaningful guarantees as well as to enable better enforcement of the correctness of data transparency disclosures. An essential ingredient for verification is the continued development of empirical techniques for data verification. Following previous work identifying key technical topics for AI governance~\citep{reuel2024open}; we give a deeper analysis of technical research directions for the verifiability of data disclosures, since significant technical bottlenecks prevent efficient auditing of the truthfulness of data transparency disclosures. Both of the following directions are particularly important for transparency objectives such as protecting personal information (Section~\ref{sec:personal-info}) and copyright protection (Section~\ref{sec:copyright}).  

\textbf{Granularity of data usage} A key challenge in data disclosure is deciding the granularity with which data membership should be reported. Granularity refers to the specificity level at which data usage is tracked and disclosed~\citep{maini2024llm}. At the coarsest granularity, developers might simply acknowledge that a dataset was used; finer granularity could involve tracking specific documents, paragraphs, or even exact text sequences. More granular disclosure introduces computational overhead but yields more meaningful transparency - for example, sequence-level reporting might require precise but potentially large lookup tables or $n$ gram statistics. Potential research directions include:
\begin{itemize}
    \item Developing compact representations of training examples that enable efficient lookups;
    \item Studying trade-offs between data privacy, utility of transparency, and computational feasibility across different levels of granularity.
\end{itemize}

\textbf{Revisiting definitions for training set inclusion} 
A more fundamental issue with training data disclosure is that data membership is an inherently fuzzy concept. Practitioners typically adopt lossy definitions and tests to operationalize fuzzy concepts, which inevitably leaves room for ambiguity or even malpractice in data transparency. Consider the recent U.S. district court ruling granting data owners permission to “inspect” developers’ training data for unauthorized content use~\cite{tremblay2023}. A non-technical inspector or auditor might simply perform a substring search—implicitly using $n$-gram overlap as a membership test—while their true intent is to identify semantically equivalent use (e.g., typos, paraphrasing, or even multilingual translations). However, ~\citeauthor{liu2025language} recently demonstrated that LLMs could synthesize and regurgitate text without including any of its original $n$-grams from the training data; LLMs effectively “stitch together” fragments due to strong generalization. In such cases, any disclosure mandates based on $n$-gram overlap may be circumvented by model developers.  
Potential research directions include:
\begin{itemize}
    \item Data membership definitions or tests that do not rely on $n$-gram overlap, or are robust to text perturbations.
    \item Probabilistic definitions of membership and as well as probabilistic testing methods.
\end{itemize}

We include two other directions: \textbf{Canary injection and data watermarks}, and \textbf{cryptographic approaches for data usage} in the supplementary materials. 

\subsection{Towards Better Data Transparency Policy}

Our analysis highlights that (1) mandated disclosures should provide enough information to be actionable for consumers, data creators, auditors, and regulators, 
(2) enforcement mechanisms for compliance and verification of these disclosures are crucial, and (3) disclosures should be designed to change the behavior of AI system developers in a way that is aligned with intended policy goals. While European policy on transparency includes enforcement mechanisms that improve compliance, whether the information required is actually informative enough or whether there is a pathway for accountable change remains to be seen. Overburdening small companies without real change from larger corporations is an outcome to be avoided~\citep{fan2022hidden}. For policymakers considering AI transparency legislation in the future, our analysis specifically illustrates the importance of the following: 
\begin{itemize}
    \item \textbf{Clarity}: Disclosures aimed at addressing too many things may cause confusion. It is more effective to focus on a clear policy objectives and ensure that the transparency measures that correspond to this goal have sufficient enforcement mechanisms. Disclosure is not the right tool to achieving every policy goal. 
    \item \textbf{Standardization}: Successful implementation depends on standardized reporting. For example, provide a common format and a website where all companies that are subject to disclosure requirements can submit their information. This approach has been adopted by the EU AI office in their new foundation model reporting guidelines~\citep{EC2025GPAIGuidelines}.
    \item \textbf{Information Intermediation}: When public resources are limited, policy should be designed to empower intermediaries, such as private litigants or third parties, capable of analyzing disclosed data.
\end{itemize}

While these recommendations are a starting point for today's AI landscape, policymakers will likely have to iteratively update transparency requirements to account for the shifting technical landscape and the actual observed impact of requiring data transparency disclosures. Ultimately, our paper emphasizes that disclosure is not a free lunch for accountability -- careful alignment of goals with objectives and robust enforcement mechanisms are necessary.


\bibliographystyle{ACM-Reference-Format}
\bibliography{aaai25}

\appendix
\newpage
\appendix
\newpage
\section{Other Objectives}
\label{app:other-objectives}
\paragraph{Analyses of Environmental Impact}
In current practice, the size of the dataset used for training provides crucial information on how much compute is used to train the model. Work in model scaling laws associates the predicted performance with the number of total flops, which is derived from the size of the dataset used for training~\citep{hoffmann2022training}. Thus, data transparency around the rough size of the dataset used to train a model could give a general estimate of the computational resources, and thus the environmental impacts of training the model. However, the size of the dataset itself does not directly reveal the amount of resources used for the development of the model. The former is because model developers train many iterations of the model and may only release the final version. 

\paragraph{Competition between LLM developers} 
Despite the increasing availability of open source model weights (e.g., Llama, DeepSeek, and Qwen models), the actual datasets used to develop these models are not transparent. Little is known about what kind of data allowed DeepSeek's R1 to possess state-of-the-art math and coding skills~\cite{guo2025deepseek}. In addition, deals for exclusive access to data~\citep{google2023reddit}, including test sets~\citep{russey2024openai}, are beginning to appear in the data supply chain for AI models. Better transparency in data collection, exclusive data contracts, and ethical data collection practices could actually develop healthy competition between LLM developers and better allow new players to enter the market with competing services. 

\paragraph{Evaluation Validity: Test Set Transparency}
Thus far, our work focuses on datasets themselves as static objects. However, with the growing popularity of inference time computation, the capabilities and behaviors of models are increasingly defined by data generated at test time. For example, given a math question, a model might generate a series of answers and rely on a verifier (e.g., another model or a proof system) to provide feedback in order to arrive at a final answer. Thus, being able to compare the performance of different models in a reliable manner may require transparency about the data that are produced at test time. 

\section{Research Directions for Computer Scientists for Training Data Disclosure}
\label{app:research-dir}
For many components of compliance and verifiability that we have discussed, technical feasibility comes into play. For instance, given a piece of data, are we able to tell whether a model has actually been trained on it?
On a technical level, training data disclosure fundamentally involves the concept of training data membership—or training set inclusion—of a set of examples~\citep{shokri2016membership}, as well as how to precisely define and verify such membership. This section discusses several topics and potential research directions related to both enabling and verifying training data disclosure. Technical research in these directions would be instrumental for understanding the kinds of policies and regulations are actually enforceable in practice. Our discussions focus primarily on modern AI systems based on large language models (LLMs), though the principles discussed may generalize to other types of AI systems and modalities.

\textbf{Granularity of data usage} A key challenge in training data disclosure is deciding the granularity at which data membership should be reported. Granularity refers to the specificity level at which data usage is tracked and disclosed~\citep{maini2024llm}. At the coarsest granularity, developers might simply acknowledge that a dataset was used; finer granularity could involve tracking specific documents, paragraphs, or even exact text sequences. More granular disclosure introduces computational overhead but yields more meaningful transparency—for instance, sequence-level reporting might require precise but potentially large lookup tables or $n$-gram statistics. Potential research directions include:
\begin{itemize}
    \item Developing compact representations of training examples that enable efficient lookups;
    \item Studying trade-offs between data privacy, utility of transparency, and computational feasibility across different levels of granularity.
\end{itemize}

\textbf{Revisiting definitions for training set inclusion} 
A more fundamental issue with training data disclosure is that data membership is an inherently fuzzy concept. Practitioners typically adopt lossy definitions and tests to operationalize fuzzy concepts, which inevitably leaves room for ambiguity or even malpractice in data transparency.

Consider the recent U.S. district court ruling granting data owners permission to “inspect” developers’ training data for unauthorized content use~\cite{tremblay2023}. A non-technical inspector or auditor might simply perform a substring search—implicitly using $n$-gram overlap as a membership test—while their true intent is to identify semantically equivalent use (e.g., typos, paraphrasing, or even multilingual translations). However, ~\citeauthor{liu2025language} recently demonstrated that LLMs could synthesize and regurgitate text without including any of its original $n$-grams from the training data; LLMs effectively “stitch together” fragments due to strong generalization. In such cases, any disclosure mandates based on $n$-gram overlap may be circumvented by model developers, allowing data use while evading inspection (see Table 21 of \cite{liu2025language} for visualization).

For the purposes of data transparency, a key for future research is to develop better membership definitions/tests and data disclosure requirements that emit lower false negatives—that is, when the disclosure reports no data usage, there is likely no data usage. Additionally, future data membership definitions and disclosure requirements should extend beyond simple set membership of text in the raw training dataset to consider data provenance, preprocessing, and other side information accessible during the model training pipeline. Potential research directions include:

\begin{itemize}
    \item Proposing data membership definitions and tests that do not rely on $n$-gram overlap, or are robust to text perturbations.
    \item Investigating probabilistic definitions of membership and as well as probabilistic testing methods.
    \item Analysis of the advantages and disadvantages of using model-based membership definitions (e.g., using model completions as a test for membership). More broadly, understanding whether data disclosure requirements be based on model behavior, rather than simply on the static training set.
\end{itemize}

\textbf{Canary injection and data watermarks} 
Canary injection involves intentionally inserting unique identifiable data points (canaries) into training datasets to verify data use in trained models~\citep{wei2024proving}. Data watermarking, on the other hand, explores injecting (human-imperceptible) statistical signals into training data that persist in model outputs, which can then aid in data usage detection~\citep{sander2024watermarking}. Unlike post hoc verification techniques such as membership inference attacks~\citep{shokri2016membership} that depend on (fuzzy) data membership definitions (see above), or self-reported metrics that are exploitable~\citep{zhang2024language, liu2025language}, canaries and watermarks allow for indisputable, side channels for data usage verification. 

The key challenges of canaries and watermarks include making them persistent (such that data usage cannot be concealed) and salient (such that data usage is easily detectable for data transparency enforcement). Potential research directions include:

\begin{itemize}
    \item Developing methods that are robust to adversarial data perturbation aiming to remove the canaries/watermarks;
    \item Developing methods against intentional or inadvertent post-hoc model sanitization; and
    \item Evaluating the effectiveness of these methods across diverse LLM architectures and training paradigms.
\end{itemize}

\textbf{Cryptographic approaches for data usage.} Cryptography also may present opportunities for strong, model-agnostic evidence about data use. Potential research directions include:
\begin{itemize}
\item Exploring methods that commit to datasets before training, and subsequently issuing publicly verifiable proofs that every optimization step consumes only data that are consistent with that commitment~\citep{jia2021proof, fang2023proof, abbaszadeh2024zero}
\item Systems research enabling modern training within trusted execution environments (TEEs)—for instance, sealing datasets within enclaves, with remote attestation certifying enclave identity and code hash before data release~\citep{intelsgx}.
\item Efficiency optimization for accelerating existing cryptographic primitives for machine learning workloads, such as fully homomorphic encryption~\citep{TFHE-rs}.
\end{itemize}
A key drawback of cryptographic approaches is their dependence on exactness. For instance, hashing training sequences would differ significantly even with minor $n$-gram variations, despite identical semantics. Therefore, defining appropriate data granularity and training set inclusion tests—as discussed previously—is crucial when considering cryptographic approaches for disclosure. This limitation may restrict cryptographic techniques primarily to the parts of the disclosable data that are otherwise unique in their representation (e.g., metadata, training timestamps).

\section{The Emergence of Data Transparency Regulation: A Case Study on AB 2013}
\label{sec:ab2013}
In practice, AI regulation, like any other legislation, is the result of a compromise between different stakeholders with different interests. Existing data transparency policies are the artifacts of arguments for data transparency and the push against it. To understand how the disclosure and remediation gap arose, we provide a closer analysis of the process that generated the California Bill AB2013. 

\subsection{Consumer Protection: A Driver for Data Transparency}
Introduced in January 2024, California State Bill 2013 initially defined terms related to artificial intelligence systems, developers, and synthetic data generation, along with mandates for disclosing certain aspects of the data supporting model development. Early revisions of the bill extended the bill's scope by widening  disclosure to include personal and consumer information, dataset modifications, dataset statistics, and outline exceptions for AI services dedicated solely to security and integrity. The Assembly Committee on Privacy and Consumer Protection emphasized that without knowledge of the data used to train AI products, Californians cannot make informed purchasing decisions. This position was grounded in existing privacy laws. The committee justified the disclosure of synthetic data by pointing out the risks of bias and "model collapse," assessing the proposed requirements as modest. At this stage, supporters of the bill underscored the importance of the bill for public transparency, awareness, and protection for Californians. 

Based on an academic report~\citep{chmielinski2024clear}, the judiciary committee saw data transparency as a way to reducing biases, addressing hallucinations and problematic outputs, and easing privacy and copyright issues. At this stage in the Senate, the California Labor Federation began to support the bill motivated by concerns that AI systems might be used to evaluate employee performance and hiring. The Federation stressed that workers should be made aware of the training data used for these decisions to prevent the non-consensual use of personal data and to tackle both implicit and explicit biases in training data.

\subsection{The Push Against Data Transparency}
Resistance to the bill emerged early from several key groups expressing doubts about the bill's technical feasibility, vague definitions of key terms, and weak protection of trade secrets and intellectual property. To address these concerns, committee amendments narrowed the definition of artificial intelligence and lowered disclosure requirements from “a description” to “a high-level summary.” As the bill reached Assembly Floor Analysis, lobbyists in opposition argued that revealing training data could hinder competition and terms like 'but not limited to' were too vague. As a result, the Senate version included more exceptions for domains where AI systems need not disclose training data (e.g., a GenAI system for operation of aircraft in the national airspace)  and relaxed data point disclosures to general ranges. In this effort to reconcile transparency and industry concerns, disclosure requirements become more vague.

In the Senate judiciary analysis phase, opposition expanded as a group of organizations raised further criticism mainly focused on two issues: the absence of a risk-level distinction (suggesting only high-risk AI should meet data transparency standards) and the broad interpretation of "artificial intelligence system or service." These concerns shaped subsequent Senate floor amendments, notably narrowing the bill's scope to "generative artificial intelligence" and restricting disclosure obligations to the original developers of a GenAI system, even if modified by a third party.

The forces for and against data transparency illustrates the path from the idea of enabling consumer protection through data transparency to passing a bill with very high-level disclosures. The end product neither ensures that the disclosures are sufficient for the original goal nor provide enforcement provisions to ensure compliance of these disclosures.  

\subsection{AB2013: Transparency In Practice}
As of the submission of this paper, AB2013 came into effect in California and the response from different companies illustrate the different challenges facing data transparency policy. 
\paragraph{Compliance}
Major providers such as OpenAI, Anthropic, and Google have all posted transparency documentation and reports of various levels of detail. OpenAI has posted a short document with high level descriptions, for example:
\begin{quote}
We develop these systems using a variety of data sources, including publicly available data, data that we partner with third parties to access, and information that our users or human trainers and researchers provide or generate. We also develop our systems using synthetic data.
\end{quote}
They further comment that they take steps to ``\textit{reduce the amount of personal information in our training datasets}'' and users have the option to opt out of their data being used for training.\footnote{\url{https://help.openai.com/en/articles/20001044-training-data-summary-pursuant-to-california-civil-code-section-3111}} Google provides a similarly brief statement that covers all of their Generative AI services from Ads to Pixel to Search to Youtube to Gemini.\footnote{url{https://transparencyreport.google.com/?hl=en}} While their descriptions are also vague, they directly admit that they might be training on copyright data and consumer information: 
\begin{quote}
Datasets may include or reference material that is subject to copyright, trademark, or patent protection, and these datasets
are used under applicable licenses or pursuant to law. Some of the datasets also include aggregate consumer information and/or personal information. 
\end{quote}
Anthropic provides training data documentation according with AB2013 with the most detail and attempts to capture all the specified requirements \footnote{\url{https://trust.anthropic.com/resources?s=km4kixlo1o4zwqj0ybiaw&name=ab-2013-training-data-summary}}. Compared to Google and OpenAI, Anthropic also offers some details around data processing and data acquisition methods. However, none of these disclosures are specific enough to guarantee protection of personal data and copyright, nor do they provide enough information for actionable change to training data practices. 

\paragraph{Legal Challenges}
In contrast, xAI, the developer of Grok, filed a lawsuit in federal court to block the statute, arguing that it violates the Fifth Amendment (for compelling disclosure of trade secrets) and the First Amendment (for forcing speech)~\citep{xai_v_bonta_2025}. The lawsuit accuses AB2013 as a ``\textit{trade-secrets-destroying disclosure regime that hands competitors a roadmap to learn how companies like xAI are developing and training their proprietary AI models}''. Beyond complaints of not being able to keep propriatary secrets, the lawsuit itself highlights that AB2013 does not make clear how this information would be valuable to consumers, thus competitors are the true beneficiaries of the bill. This echos the specification gap we describe: the existence of disclosures does not necessarily imply achieving the originally desired transparency goal.

\end{document}